\documentclass[a4paper,11pt]{article}
\usepackage{pos}

\usepackage{epstopdf}
\newcommand{\Pom}{\mathbb{P}}

\newcommand{\Reg}{\mathbb{R}}

\newcommand{\bq}{\mbox{\boldmath $q$}}
\newcommand{\bk}{\mbox{\boldmath $k$}}

\renewcommand\slash[1]{\not \! #1}

\newcommand{\p}{\partial}

\newcommand{\twosidep}[1]{\stackrel{\leftrightarrow}{\p}_{\! #1}}
\newcommand{\leftsidep}[1]{\stackrel{\leftarrow}{\p}_{\! #1}}
\newcommand{\rightsidep}[1]{\stackrel{\rightarrow}{\p}_{\! #1}}

\title{Central exclusive production of $\eta$ and $\eta'$ mesons in diffractive proton-proton collisions at the LHC and the nature of the pomeron}
\ShortTitle{Central exclusive production of $\eta$ and $\eta'$ mesons in diffractive $pp$ collisions}

\author[a]{Piotr Lebiedowicz}
\author*[b]{Otto Nachtmann}
\author[a,c]{Antoni Szczurek}

\affiliation[a]{Institute of Nuclear Physics Polish Academy of Sciences,\\ Radzikowskiego 152, PL-31342 Krak{\'o}w, Poland}

\affiliation[b]{Institut f\"ur Theoretische Physik, Universit\"at Heidelberg,\\ Philosophenweg 16, D-69120 Heidelberg, Germany}

\affiliation[c]{Institute of Physics, Faculty of Exact and Technical Sciences, University of Rzesz{\'o}w, \\
Pigonia 1, PL-35310 Rzesz{\'o}w, Poland}

\emailAdd{Piotr.Lebiedowicz@ifj.edu.pl}
\emailAdd{O.Nachtmann@thphys.uni-heidelberg.de}
\emailAdd{Antoni.Szczurek@ifj.edu.pl}

\abstract{Central exclusive production (CEP) of $\eta$ and $\eta'$ mesons
in proton-proton collisions at high-energies is discussed.
At the LHC the main mechanism for the production of these mesons
should be double-pomeron ($\Pom$-$\Pom$) exchange,
that is, the fusion reactions $\Pom \Pom \to \eta, \eta'$.
We show that for a scalar pomeron these fusion reactions
are not possible. In contrast, in the tensor-pomeron model
CEP of $\eta$ and $\eta'$ mesons via double-pomeron exchange
is allowed.
We discuss these reactions for the c.m. energy $\sqrt{s} = 29.1$ GeV
realised at the WA102 experiment and for $\sqrt{s} = 13$ TeV 
corresponding to the LHC experiments.
Cross sections and distributions are presented and discussed.}

\FullConference{Proceedings of the 
53rd International Symposium on Multiparticle Dynamics (ISMD 2025)\\
21 - 26 September, 2025\\
Corfu, Greece\\}

\begin{document}
\maketitle

\section{Introduction}
\label{section:1}

We want to discuss central exclusive production (CEP) 
in proton-proton collisions at high energies, in particular,
at the LHC,
\begin{eqnarray}
p(p_{a}) + p(p_{b}) \to p(p_{1}) + X + p(p_{2}) \,.
\label{1.1}
\end{eqnarray}
We require large rapidity gaps between $X$ and the outgoing protons.
For $X$ we have many options:
\begin{eqnarray}
X: \eta,\, \eta',\, f_{0},\, f_{1},\, f_{2},\, f_{4},\, 
\rho^{0},\, \omega,\, \phi,\, 
\pi^{+}\pi^{-},\,
K^{+}K^{-},\, p \bar{p},\, 
\phi \phi,\, \rho^{0}\rho^{0}, \, K^{*0} \bar{K}^{*0},\, 
4\pi,\, 4K, \ldots
\label{1.2}
\end{eqnarray}
In our group we have written quite a number of papers on this subject
\cite{Lebiedowicz:2013ika,Lebiedowicz:2014bea,Lebiedowicz:2016ioh,Lebiedowicz:2016zka,Lebiedowicz:2019por,Lebiedowicz:2019boz,Lebiedowicz:2019jru,Lebiedowicz:2020yre,Lebiedowicz:2021pzd,Lebiedowicz:2025num,Lebiedowicz:2025xob}.
Here we shall concentrate on $\eta$ and $\eta'$ CEP 
which we treated in \cite{Lebiedowicz:2025num}.

\begin{figure}[!ht]
\centering
\includegraphics[width=5.5cm]{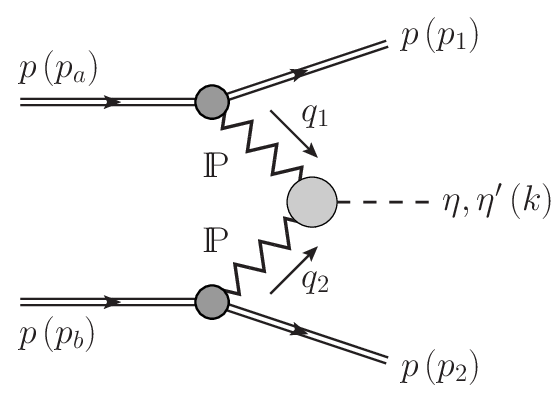}
\caption{\label{fig:1}
CEP of $\eta$ and $\eta'(958)$ with double-pomeron exchange.
The pomeron momenta are $q_{1}$, $q_{2}$, and $k = q_{1} + q_{2}$
is the momentum of the produced meson.}
\end{figure}

At high energies these reactions should be due to pomeron-pomeron fusion (Fig.~\ref{fig:1}).
These processes are very interesting since they can tell us
something about the \underline{nature of the} \underline{pomeron}.

If $\eta$ and $\eta'$ CEP is observed at the LHC a scalar pomeron is excluded.
To see this let us look at Fig.~\ref{fig:1}, e.g. for $\eta'$ CEP.
At the center we have, assuming a scalar pomeron $\Pom_{\rm S}$,
the vertex
\begin{eqnarray}
\Gamma^{(\Pom_{\rm S} \Pom_{\rm S} \to \eta')}(q_{1},q_{2})\,.
\label{1.3}
\end{eqnarray}
From Lorentz invariance this vertex can only depend on 
$q_{1}^{2}$, $q_{2}^{2}$, and $k^{2} = (q_{1} + q_{2})^{2}$:
\begin{eqnarray}
\Gamma^{(\Pom_{\rm S} \Pom_{\rm S} \to \eta')}(q_{1},q_{2})
=
F(q_{1}^{2},q_{2}^{2},k^{2})\,.
\label{1.4}
\end{eqnarray}
From parity invariance of QCD we get for the pseudoscalar $\eta'$ meson
\begin{eqnarray}
\Gamma^{(\Pom_{\rm S} \Pom_{\rm S} \to \eta')}(q_{1},q_{2})
=
-\Gamma^{(\Pom_{\rm S} \Pom_{\rm S} \to \eta')}(q_{1}',q_{2}')\,,
\label{1.5}
\end{eqnarray}
where
\begin{eqnarray}
q_{i}' = 
\begin{pmatrix}
q_{i}^{0}\\
-\bq_{i}
\end{pmatrix}\,,
\quad 
k' = 
\begin{pmatrix}
k^{0}\\
-\bk
\end{pmatrix}\,, \quad
i = 1,2.
\label{1.6}
\end{eqnarray}
But from (\ref{1.4})--(\ref{1.6}) we find
\begin{align}
F(q_{1}^{2},q_{2}^{2},k^{2})
&=-F(q_{1}'^{2},q_{2}'^{2},k'^{2}) \nonumber \\
&=-F(q_{1}^{2},q_{2}^{2},k^{2}) = 0\,,
\label{1.7}
\end{align}
since
\begin{eqnarray}
q_{i}'^{2} = q_{i}^{2}\,, \quad k'^{2} = k^{2}\,.
\end{eqnarray}

With a scalar pomeron CEP of $\eta$, $\eta'$ and, using a similar argument,
of $f_{1}$ is not possible.

In the paper \cite{Ewerz:2016onn} it was shown that a scalar pomeron
is not compatible with the measurements by STAR \cite{Adamczyk:2012kn}
concerning the helicity structure of $pp$ elastic scattering.
But there one is dealing with the $\Pom pp$ vertex:
one pomeron, two hadrons.
In CEP we are dealing with the $\Pom \Pom \eta$ and $\Pom \Pom \eta'$ vertices:
two pomerons, one hadron.
Thus, CEP gives independent, new, information.

\section{The tensor pomeron and CEP}
\label{section:2}

We believe that the best effective description of the pomeron
is to assume that it couples to hadrons like a rank~2 symmetric tensor;
see \cite{Ewerz:2013kda}.

Let $\Pom_{\mu \nu}(x)$ be the effective field of the tensor pomeron
and $\tilde{\chi}(x)$ the field of the pseudoscalar meson,
$\eta$ or $\eta'$.
Then we can construct two coupling Lagrangians $\Pom \Pom \tilde{\chi}$;
see (2.3)--(2.6) of \cite{Lebiedowicz:2013ika}.

\begin{align}
{\cal L}_{\Pom \Pom \tilde{\chi}}'(x) &= 
-\dfrac{2}{M_{0}} \, g_{\Pom \Pom \tilde{\chi}}' \, 
\left[ \partial_{\rho} \Pom_{\mu \nu}(x) \right] \, 
\left[ \partial_{\sigma} \Pom_{\kappa \lambda}(x) \right] \, 
g^{\mu \kappa} \, \varepsilon^{\nu \lambda \rho \sigma} \,
\tilde\chi(x) \,,
\label{2.1}\\
{\cal L}_{\Pom \Pom \tilde{\chi}}''(x) &=
-\dfrac{g_{\Pom \Pom \tilde{\chi}}''}{M_{0}^{3}} \,
\varepsilon^{\mu_{1} \mu_{2} \nu_{1} \nu_{2}} \, ( \partial_{\mu_{1}} \tilde\chi(x) )
\nonumber \\
&\quad  \times  
\left[
\left(
\partial_{\mu_{3}} \Pom_{\mu_{4} \nu_{1}}(x) -
        \partial_{\mu_{4}} \Pom_{\mu_{3} \nu_{1}}(x) \right) 
\twosidep{\mu_{2}} 
\left( \partial^{\mu_{3}} \Pom^{\mu_{4}}_{\quad \nu_{2}}(x) -
       \partial^{\mu_{4}} \Pom^{\mu_{3}}_{\quad \nu_{2}}(x) \right) 
\right],\quad
\label{2.2}
\end{align}
where the asymmetric derivative has the form
$\twosidep{\mu} \,= \,\rightsidep{\mu} - \leftsidep{\mu}$.
Here $M_{0} \equiv 1$~GeV and $g_{\Pom \Pom \tilde{\chi}}'$
and $g_{\Pom \Pom \tilde{\chi}}''$ are dimensionless coupling constants,
to be determined by comparison of theory and experiment.
From these Lagrangians we get easily the corresponding
bare vertex functions for $\eta$ and $\eta'$ CEP 
($\tilde{\chi} = \eta, \eta'$)
\begin{eqnarray}
\Gamma_{\mu \nu,\kappa \lambda}'^{(\Pom \Pom \to \tilde{\chi})}(q_{1},q_{2}) \mid_{\rm bare}\,
+ \,
\Gamma_{\mu \nu,\kappa \lambda}''^{(\Pom \Pom \to \tilde{\chi})}(q_{1},q_{2}) \mid_{\rm bare} \,.
\label{2.3}
\end{eqnarray}
%

The effective pomeron propagators contain the usual Regge factors:
\begin{align}
\includegraphics[width=120pt]{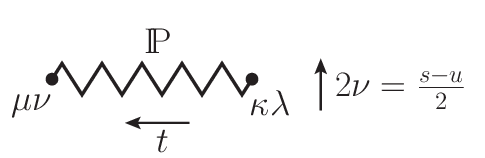} 
i \Delta^{(\Pom)}_{\mu \nu, \kappa \lambda}(2 \nu,t) = 
\frac{1}{8\nu} 
\left( g_{\mu \kappa} g_{\nu \lambda} 
     + g_{\mu \lambda} g_{\nu \kappa}
     - \frac{1}{2} g_{\mu \nu} g_{\kappa \lambda} \right)
(-i \,2 \nu \,\alpha'_{\Pom})^{\alpha_{\Pom}(t)-1}\,.
\label{2.4}
\end{align}
The pomeron trajectory $\alpha_{\Pom}(t)$ is
assumed to be of the standard form:
\begin{eqnarray}
\alpha_{\Pom}(t) = 1 + \epsilon_{\Pom} +\alpha'_{\Pom} t\,,
\quad 
\epsilon_{\Pom} = 0.0808\,, 
\quad
\alpha'_{\Pom} = 0.25 \; \mathrm{GeV}^{-2}\,.
\label{2.5}
\end{eqnarray}
At high c.m. energies $\sqrt{s}$ and small $|t|$ we have $2 \nu \to s$.
For the $\Pom pp$ vertex we follow the ansatz given in \cite{Ewerz:2013kda}:
\begin{eqnarray}
i\Gamma_{\mu \nu}^{(\Pom pp)}(p',p)
=-i 3 \beta_{\Pom NN} F_{1}[(p'-p)^{2}]
\left[ 
\frac{1}{2} \gamma_{\mu}(p'+p)_{\nu} 
+ \frac{1}{2} \gamma_{\nu}(p'+p)_{\mu} 
- \frac{1}{4} g_{\mu \nu} (\slash{p}' + \slash{p})
\right]\,.
\label{2.6}
\end{eqnarray}
Here 
\begin{eqnarray}
\beta_{\Pom NN} = 1.87~{\rm GeV}^{-1}
\label{2.7}
\end{eqnarray}
and $F_{1}(t)$
is a form factor normalised to $F_{1}(0) = 1$.

So far we discussed the \underline{bare} $\Pom \Pom \eta$
and $\Pom \Pom \eta'$ vertices. In the actual calculations
we introduce also form factors.
We consider in addition $f_{2 \Reg}$ exchange which 
is a non-leading Regge exchange 
and we include absorptive corrections.

\section{Results}
\label{results}

\subsection{Comparison with the WA102 data}

\begin{figure}[!ht]
\centering
(a)\includegraphics[width=0.4\textwidth]{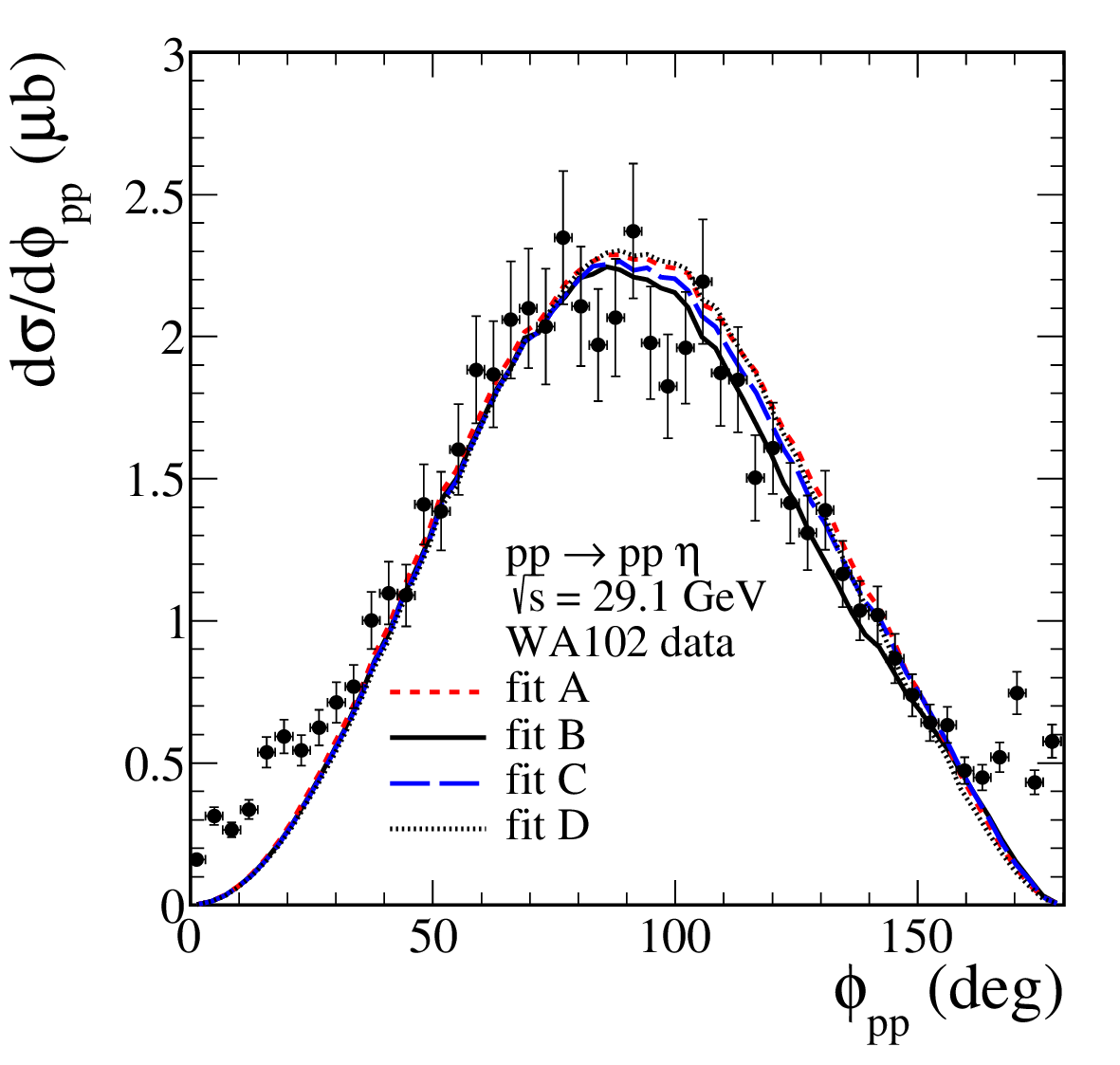}
(b)\includegraphics[width=0.4\textwidth]{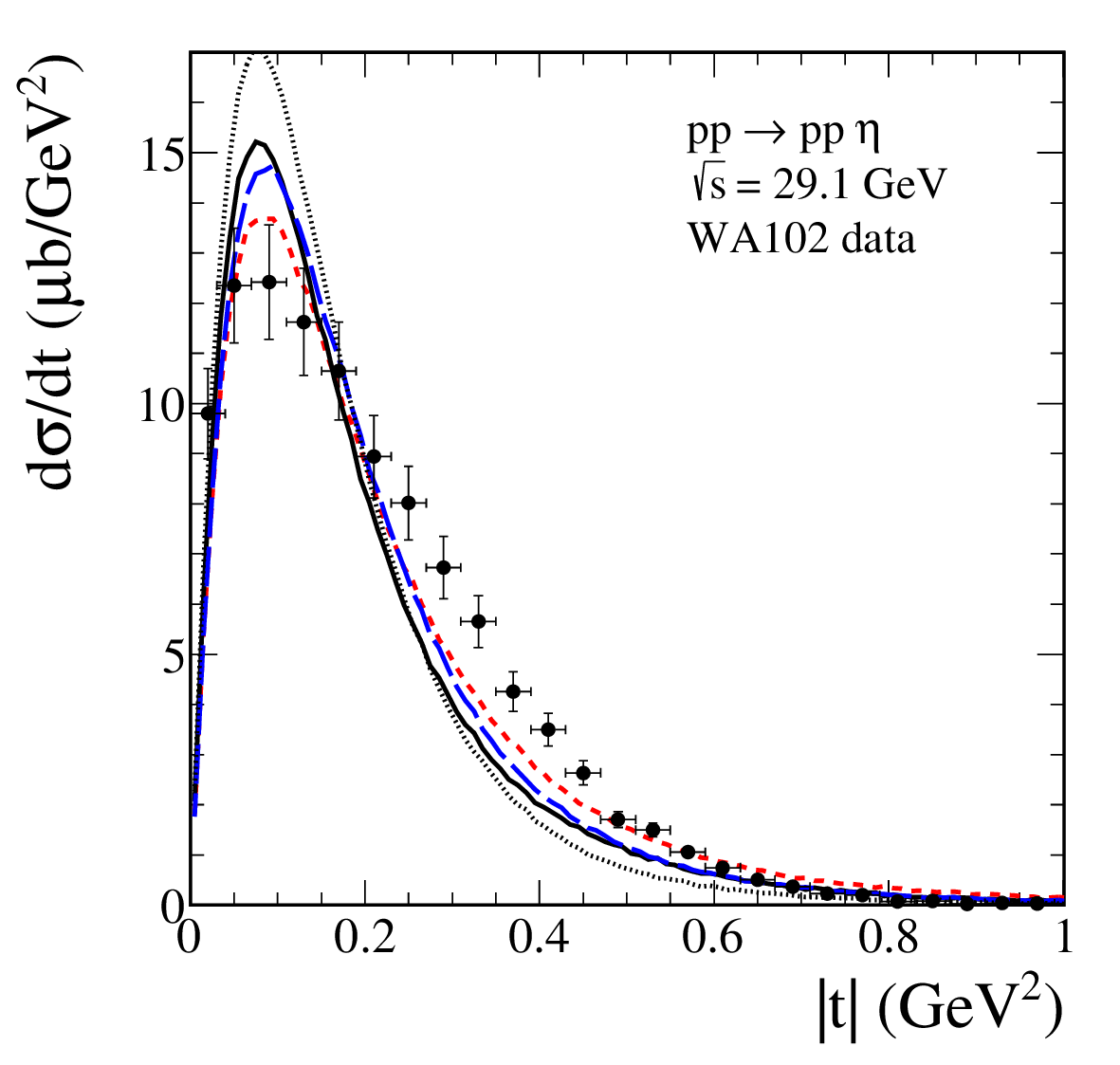}
  \caption{\label{fig:2}
Data for CEP of $\eta$ mesons [$X = \eta$ in (\ref{1.1})]
from the WA102 experiment
compared to fits from the tensor-pomeron model.
The WA102 data points from \cite{WA102:1998ixr}
have been normalised to the mean value of the total cross section
(\ref{3.1}).
(a) 
Distribution of $\phi_{pp}$, the angle between the transverse
momenta of the outgoing protons in (\ref{1.1}).
(b) 
Distribution in $t = (p_{a} - p_{1})^{2}$ or $(p_{b} - p_{2})^{2}$ 
in (\ref{1.1}). (From Fig.~3 of \cite{Lebiedowicz:2025num}).}
\end{figure}

\begin{figure}[!ht]
\centering
(a)\includegraphics[width=0.4\textwidth]{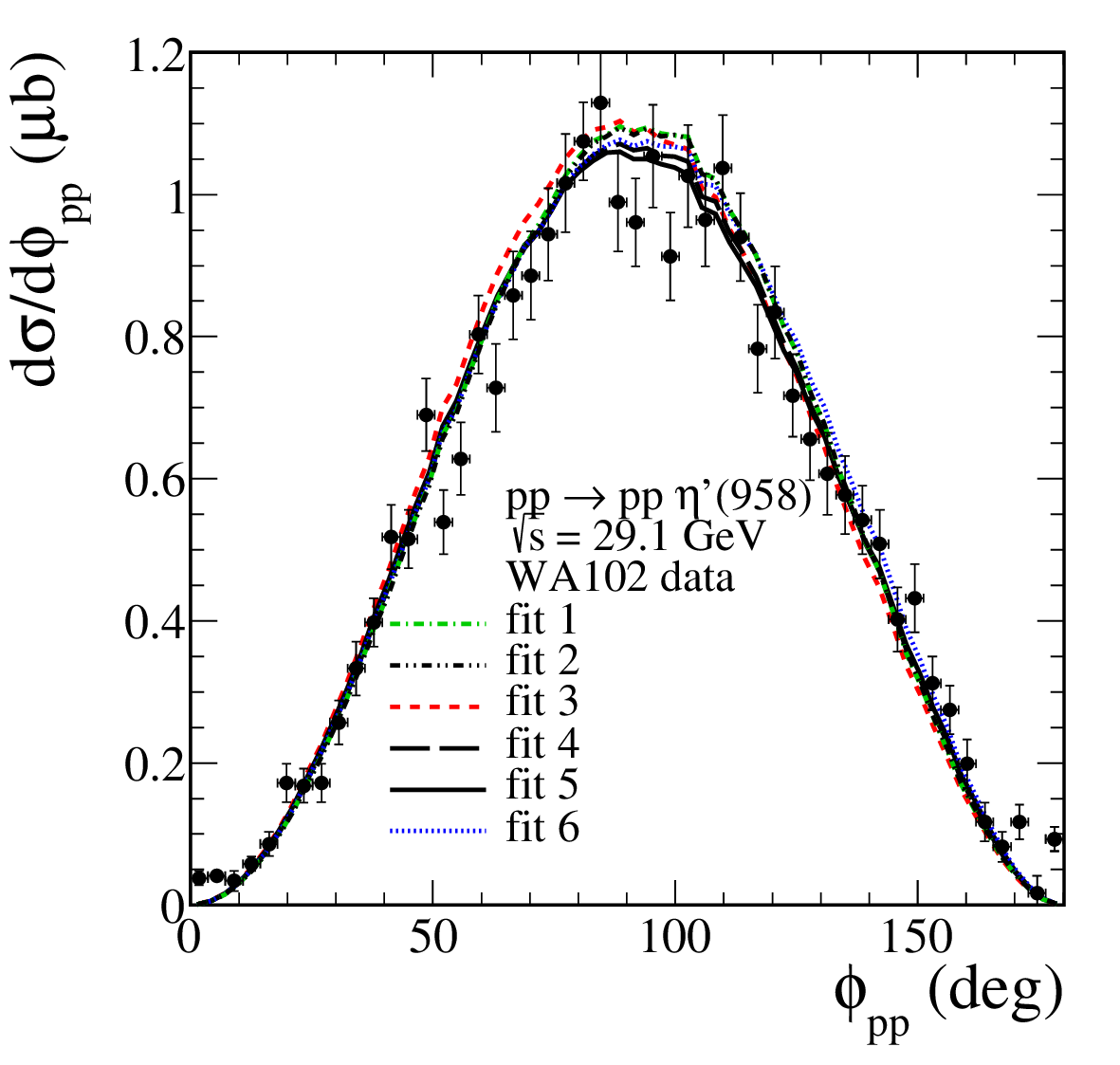}
(b)\includegraphics[width=0.4\textwidth]{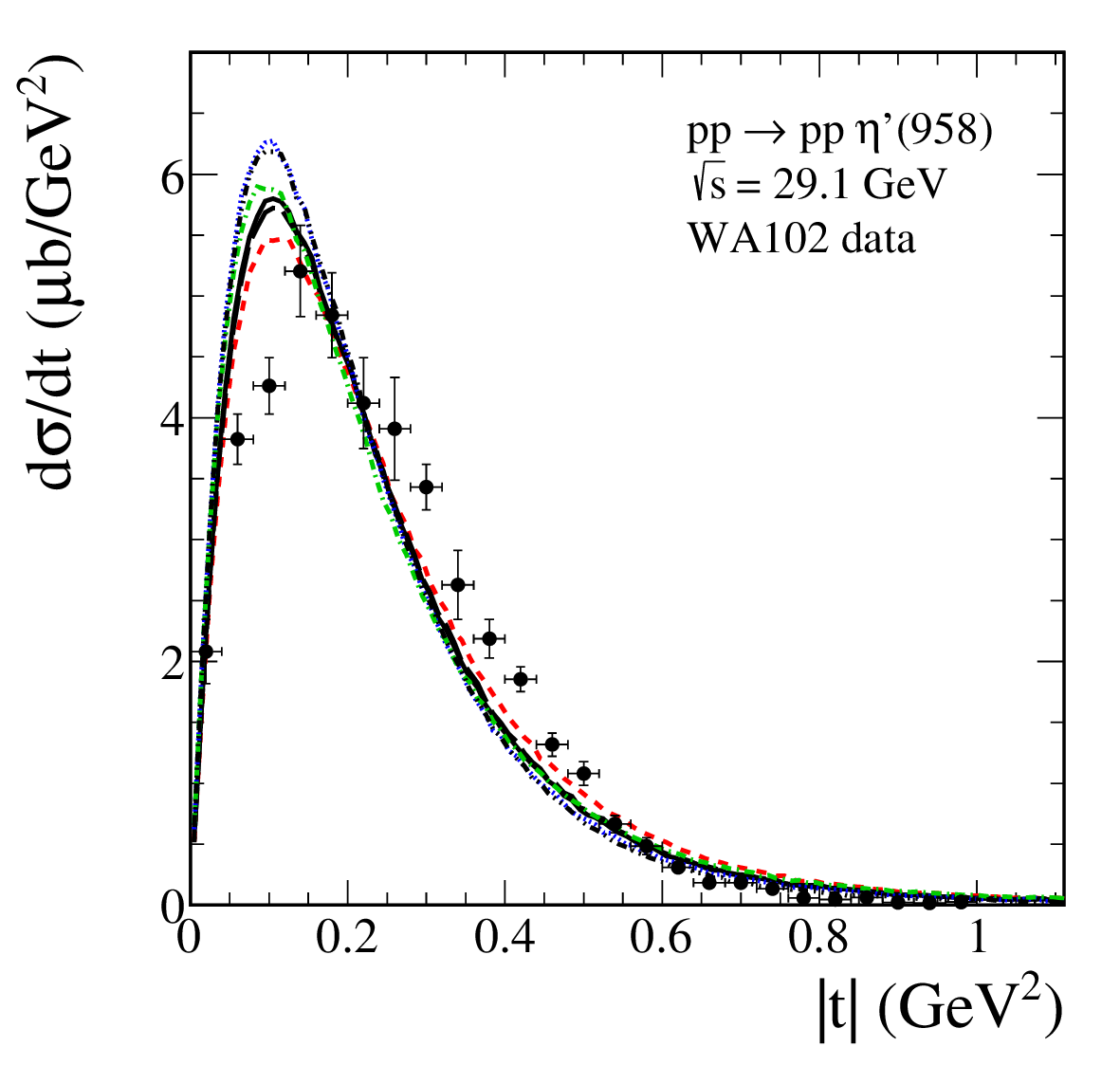}
  \caption{\label{fig:3}
Same as Fig.~\ref{fig:2} but for $\eta'$ CEP.
(From Fig.~2 of \cite{Lebiedowicz:2025num}).}
\end{figure}

There exist measurements of $\eta$ and $\eta'(958)$ CEP 
from the WA102 experiment at c.m. energy $\sqrt{s} = 29.1$~GeV.
The total cross sections are \cite{Kirk:2000ws}:
\begin{eqnarray}
\eta:  && \sigma_{\rm{exp.}} = (3859 \pm 368)~{\rm nb}\,, \nonumber \\
\eta': && \sigma_{\rm{exp.}} = (1717 \pm 184)~{\rm nb}\,.
\label{3.1}
\end{eqnarray}
In the WA102 experiment also many distributions were measured.
As examples we show in Fig.~\ref{fig:2} for $\eta$
and in Fig.~\ref{fig:3} for $\eta'$ CEP
the distributions in $\phi_{pp}$,
the angle between the transverse momenta of the outgoing protons
$p(p_{1})$ and $p(p_{2})$,
and the $t$-dependences.
In order to describe the $\phi_{pp}$ distributions
with the maximum at $\phi_{pp} \simeq \pi/2$ we need
both couplings $g'$ of (\ref{2.1}) and $g''$ of (\ref{2.2}).
In the various fits the relative contributions of pomeron
and non leading exchanges are varied. We found that the WA102 energy
is too low for pinning down the pomeron contribution.

\subsection{Predictions for the LHC experiments}

\begin{figure}[!ht]
\centering
(a)\includegraphics[width=0.4\textwidth]{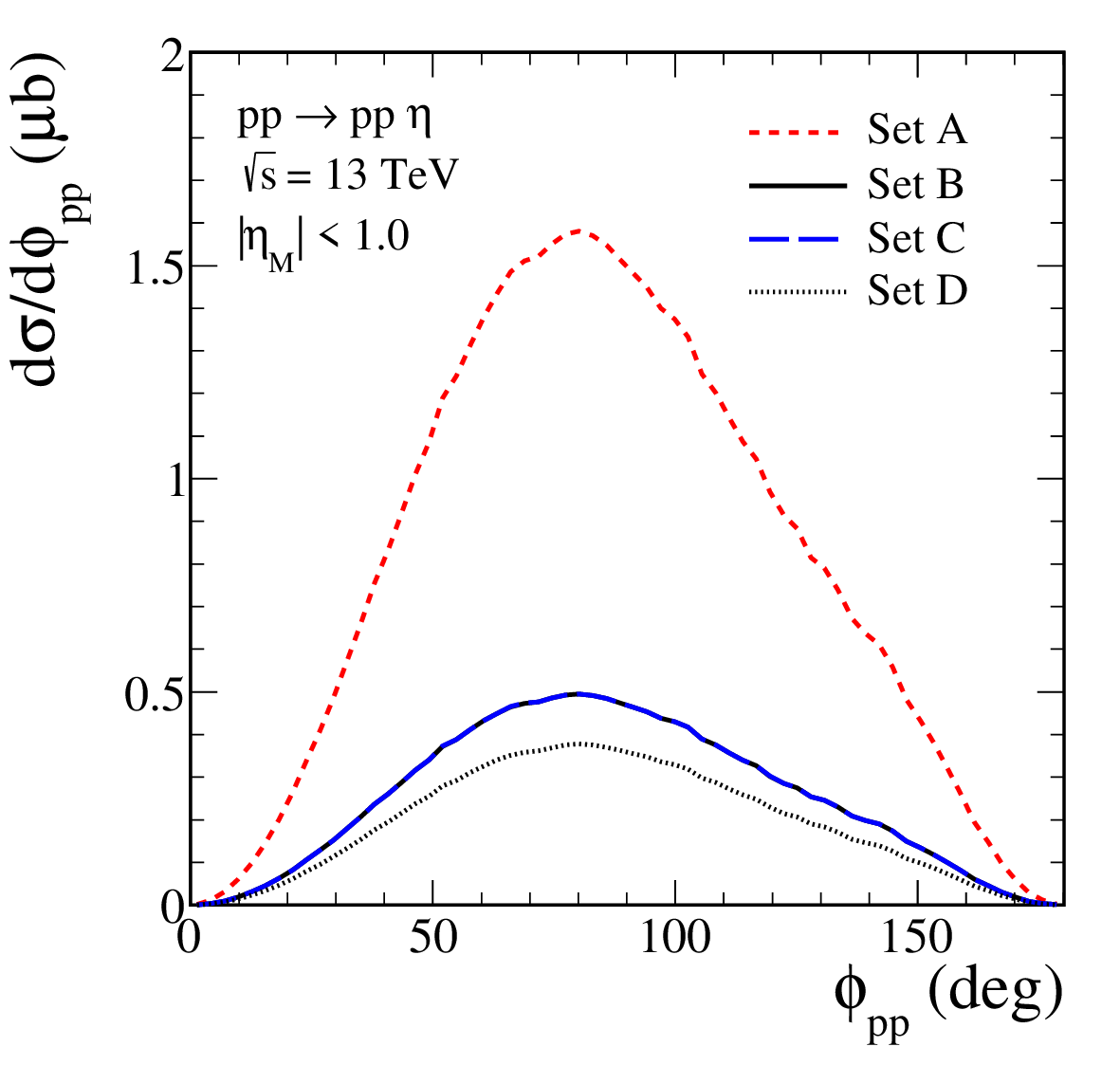}
(b)\includegraphics[width=0.4\textwidth]{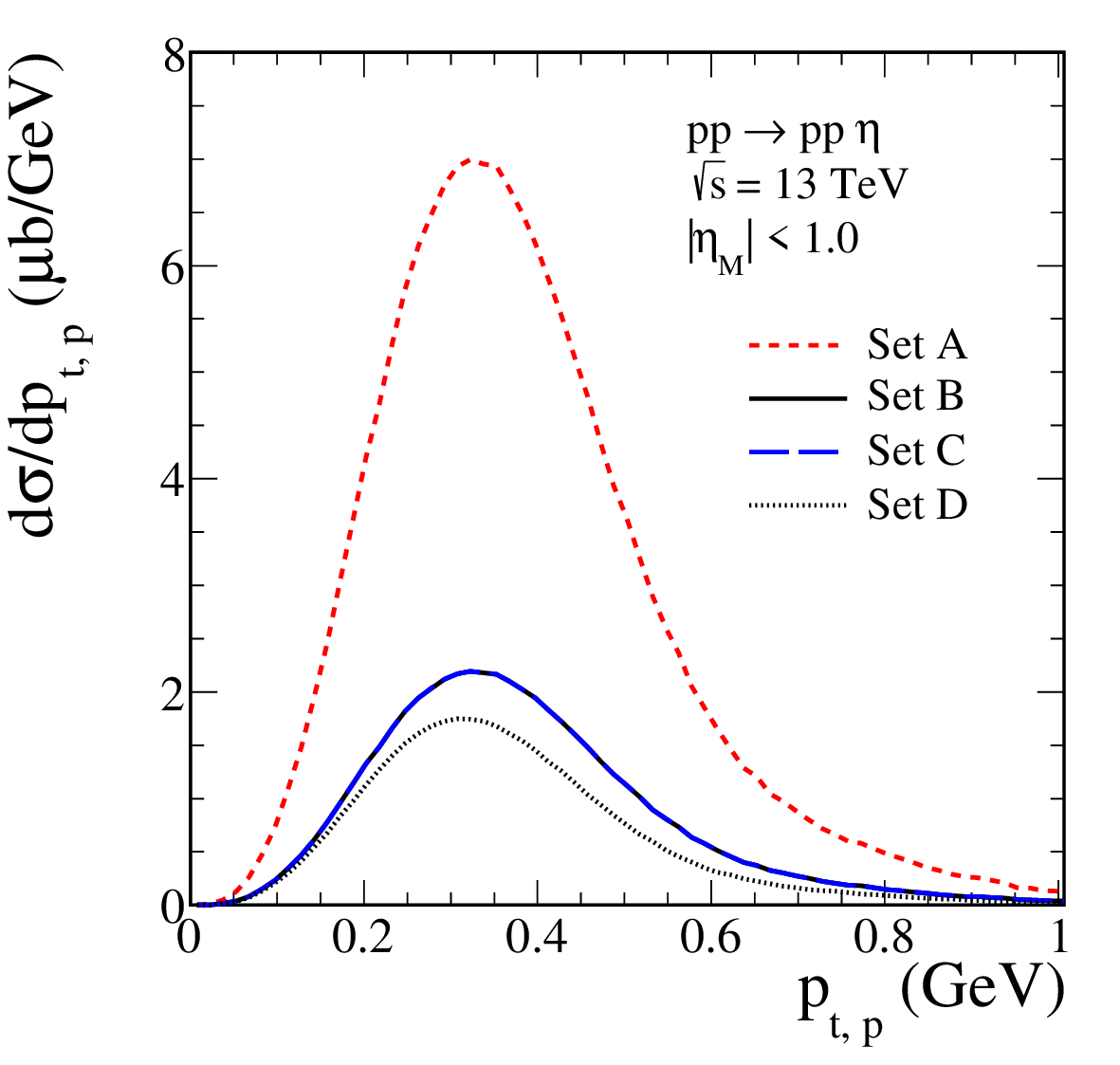}
  \caption{\label{fig:4}
Predictions for CEP of the $\eta$ meson 
calculated at c.m. energy $\sqrt{s} = 13$~TeV and for $|\eta_{M}| < 1.0$
($M = \eta$):
(a) $\phi_{pp}$ distributions,
(b) $p_{t,p}$ distributions.
(From Fig.~7 of \cite{Lebiedowicz:2025num}).}
\end{figure}

\begin{figure}[!ht]
\centering
(a)\includegraphics[width=0.4\textwidth]{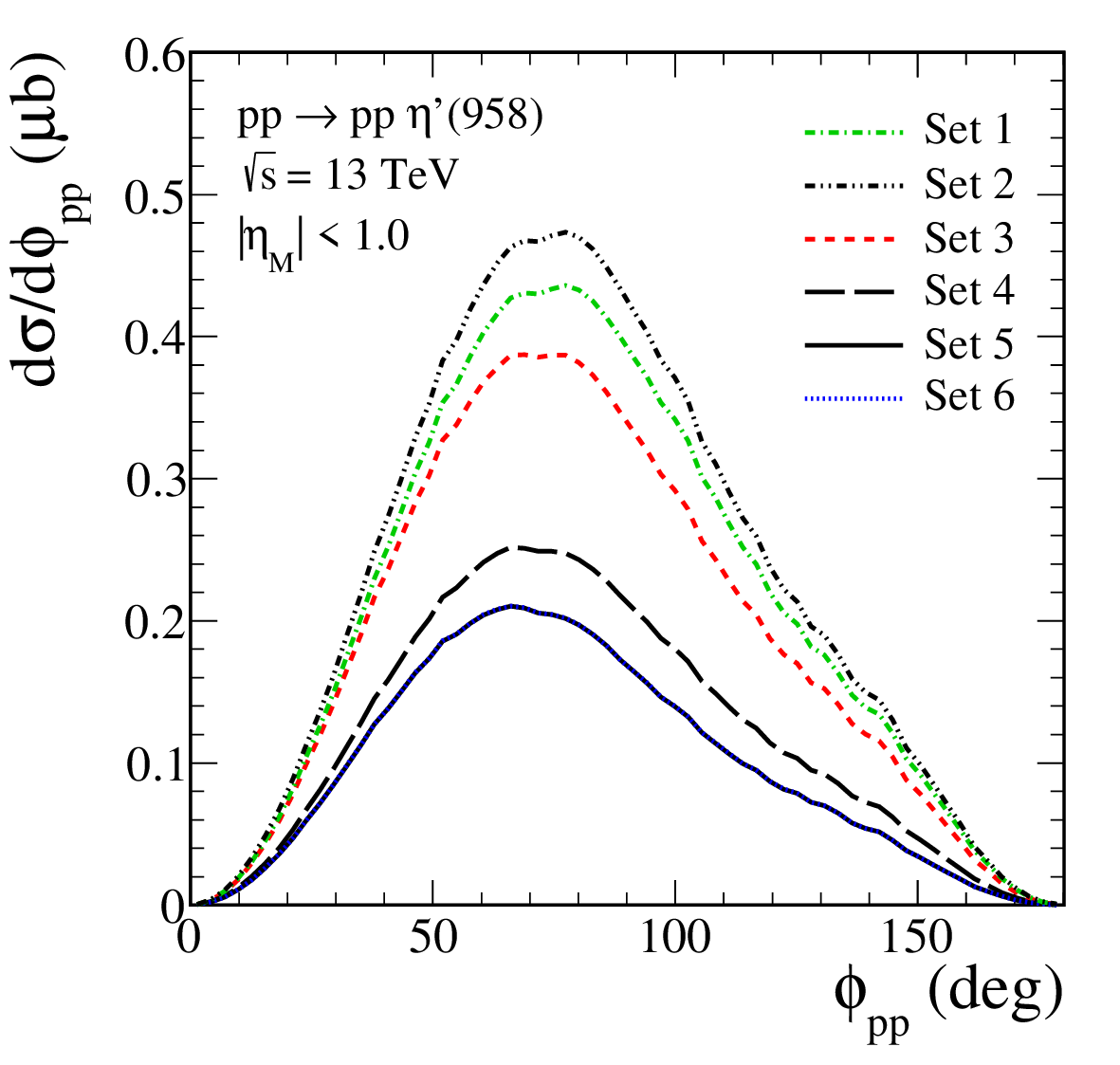}
(b)\includegraphics[width=0.4\textwidth]{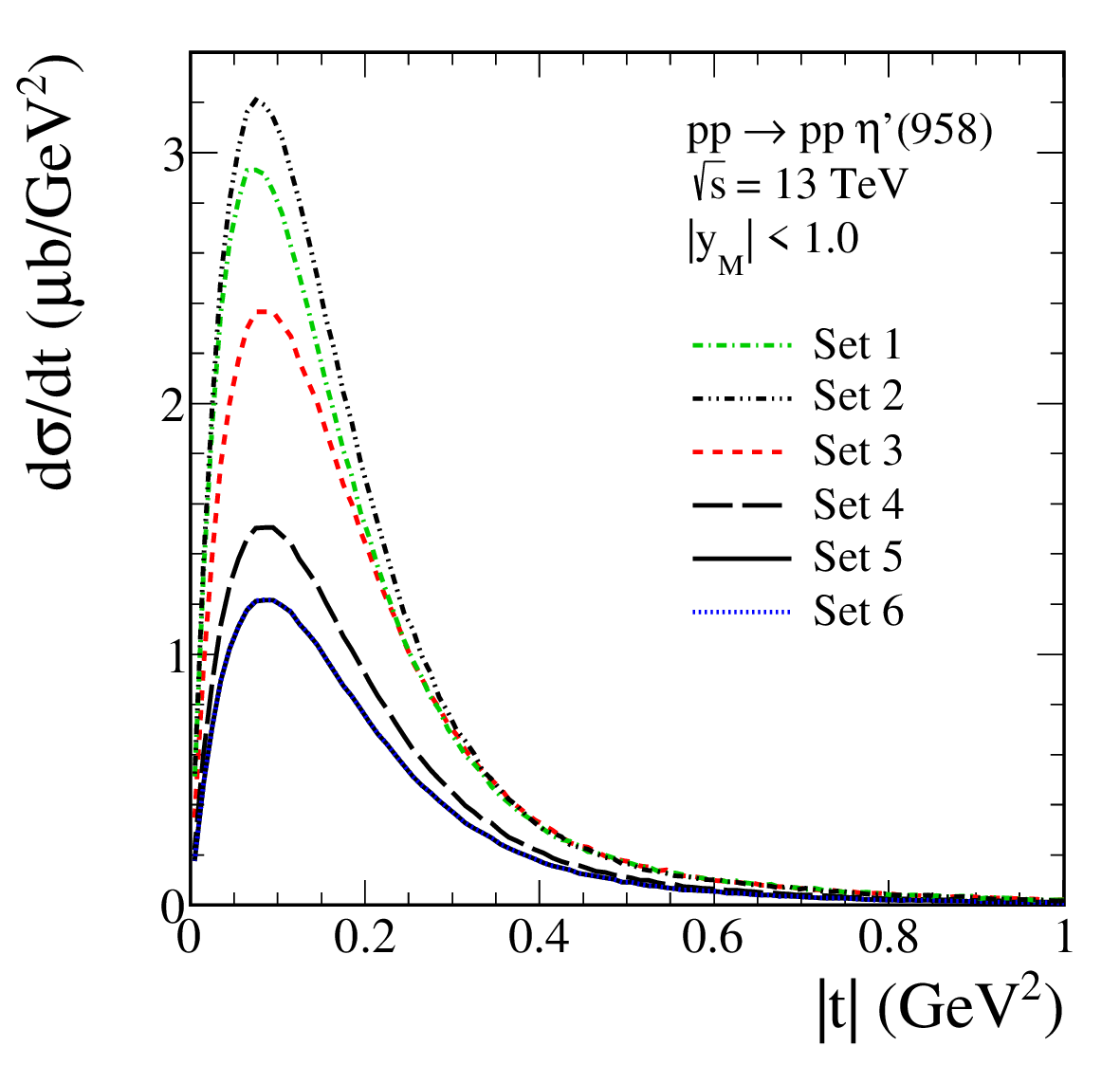}
(c)\includegraphics[width=0.4\textwidth]{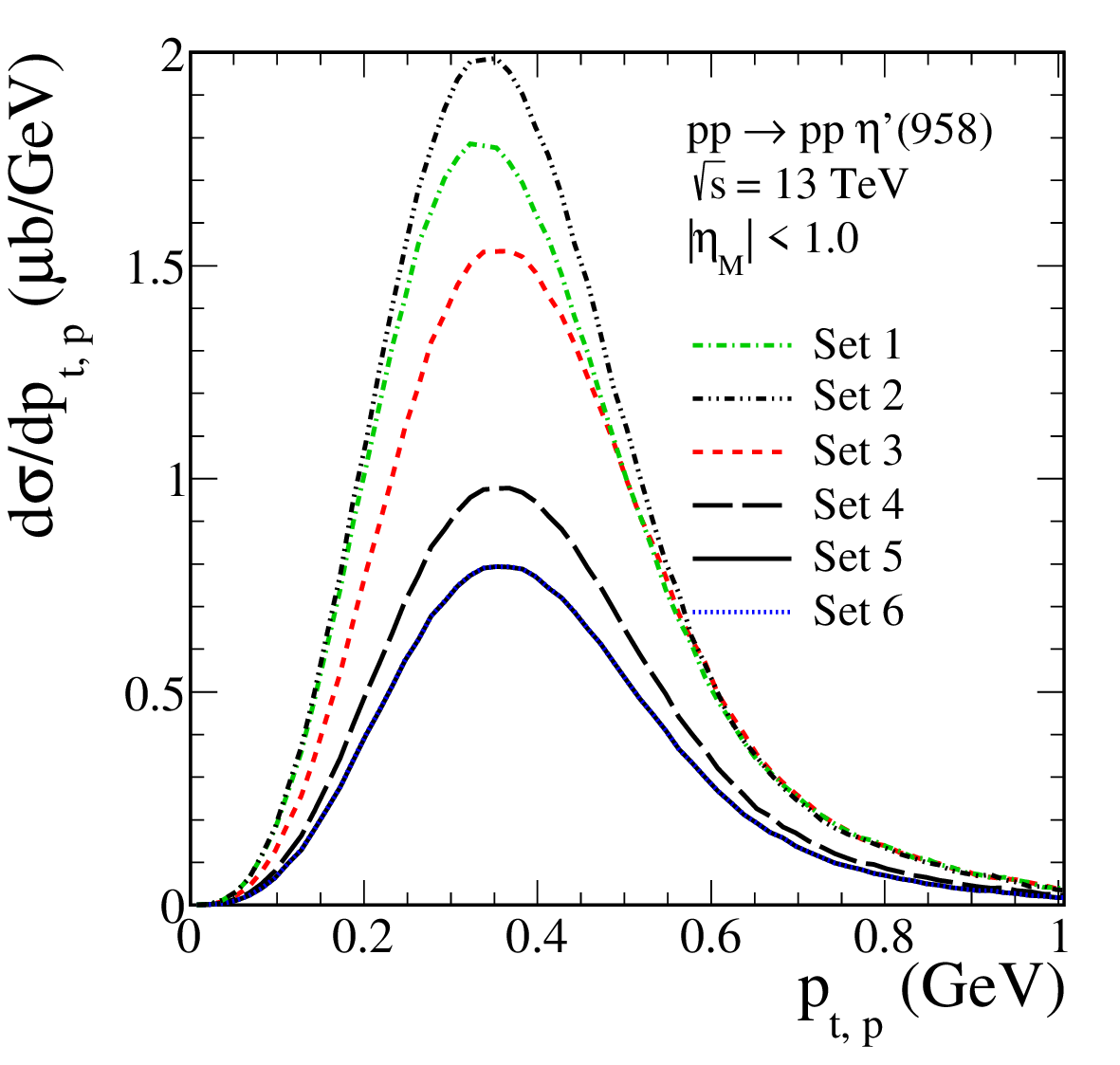}
(d)\includegraphics[width=0.4\textwidth]{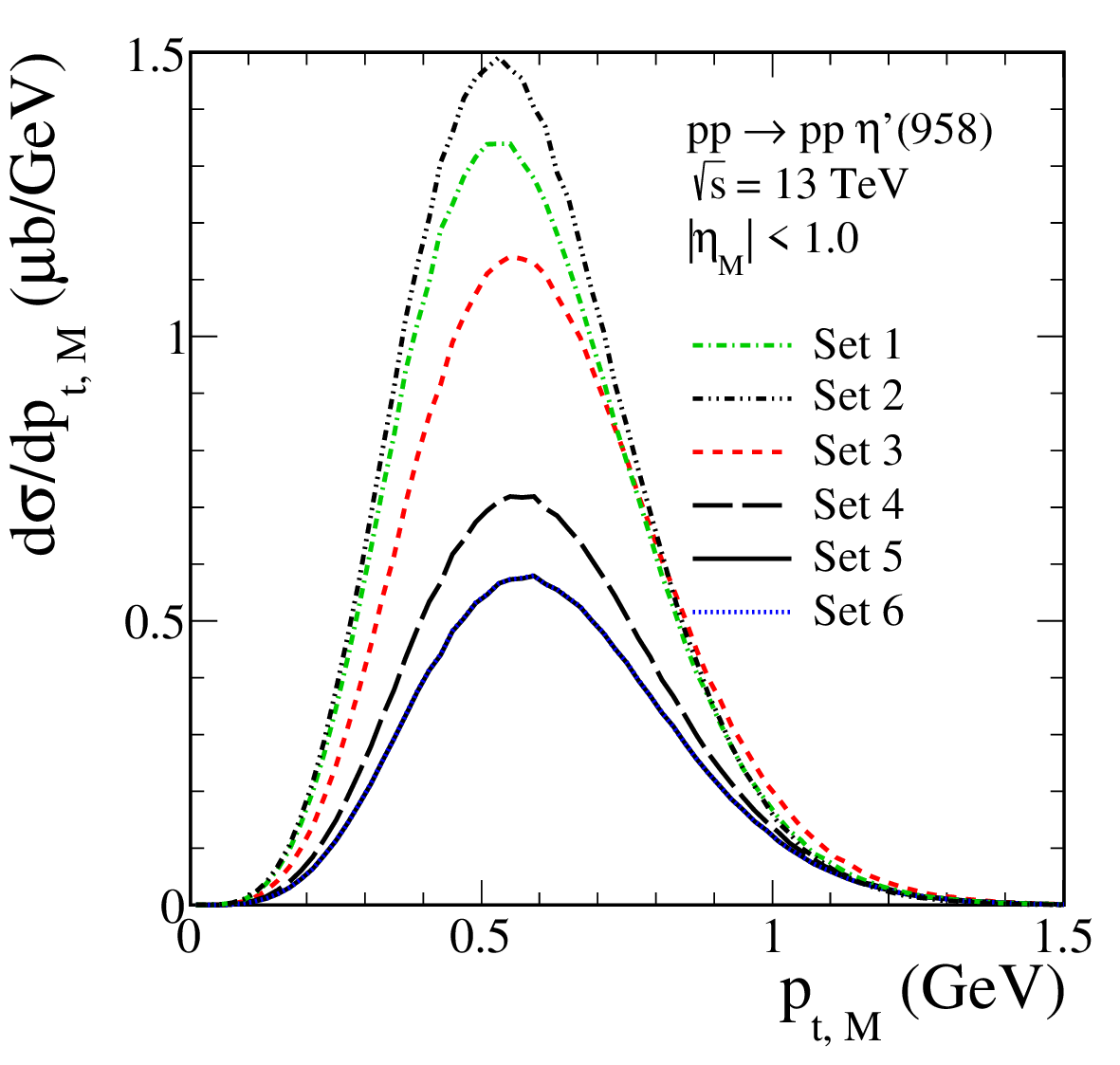}
  \caption{\label{fig:5}
Predictions for $\eta'$ CEP 
calculated at $\sqrt{s} = 13$~TeV and for $|\eta_{M}| < 1.0$.
Here $M = \eta'$.
The following distributions are shown:
(a) $\phi_{pp}$,
(b) $|t|$,
(c) $p_{t,p}$,
(d) $p_{t,M}$.
(From Fig.~5 of \cite{Lebiedowicz:2025num}).}
\end{figure}

Now we come to our predictions for the LHC experiments
where subleading $f_{2 \Reg}$ contributions should be very small.
We choose here $\sqrt{s} = 13$~TeV.
In Fig.~\ref{fig:4} and Fig.~\ref{fig:5} we show our predictions
for $\eta$ and $\eta'$ CEP where we use the following variables:
$\eta_{M}$ pseudorapidity of $\eta$ or $\eta'$ meson,
$p_{t,M}$ absolute value of the transverse momentum of $\eta$ or $\eta'$ meson,
$p_{t,p}$ absolute value of the transverse momentum of the protons
$p(p_{1})$ or $p(p_{2})$,
$\phi_{pp}$ angle between the transverse momenta of the outgoing protons
$p(p_{1})$ and $p(p_{2})$, and 
$t = (p_{a} - p_{1})^{2}$ or $(p_{b} - p_{2})^{2}$.
Our fits which were practically indistinguishable at the WA102 energy give clearly separated predictions
for LHC energies (see Table~\ref{table:1}).
\begin{table}[!h]
\centering
\caption{The integrated cross sections in $\mu$b for CEP of $\eta$
and $\eta'$ in $pp$ collisions for $\sqrt{s}=13$~TeV
for some kinematical cuts on the pseudorapidity of the mesons.
The results with absorption effects are presented.
In the last column
$S^{2} = \sigma_{\rm abs}/\sigma_{\rm Born}$ is
the ratio of the cross sections 
with ($\sigma_{\rm abs}$) and without ($\sigma_{\rm Born}$) absorption effects.
In the calculations, 
we used the parameter sets corresponding 
to the fits A--D for $\eta$ 
and to the fits 1--6 for $\eta'$;
see Figs.~\ref{fig:4} and \ref{fig:5}, respectively.
(From Table~II of \cite{Lebiedowicz:2025num}).}
\label{table:1}
\begin{tabular}{c|c|c|c|c}
\hline
\hline
Meson $M$ 
& Cuts
& Parameter set 
& $\sigma_{\rm abs}$ ($\mu$b) 
& $S^{2}$\\
\hline
$\eta'(958)$ 
& $|\eta_{M}|<1.0$ 
 & 1 &  0.66 & 0.40\\
&& 2 &  0.72 & 0.42\\
&& 3 &  0.59 & 0.39\\
&& 4 &  0.37 & 0.40\\
&& 5 &  0.30 & 0.42\\
&& 6 &  0.31 & 0.42\\
\hline
& $2.0 < \eta_{M} < 5.0$ 
 & 1 &  1.94 & 0.40\\
&& 2 &  2.09 & 0.42\\
&& 3 &  1.67 & 0.39\\
&& 4 &  1.08 & 0.40\\
&& 5 &  0.88 & 0.42\\
&& 6 &  0.88 & 0.42\\
\hline
$\eta$ 
& $|\eta_{M}|<1.0$ 
 & A &  2.51 & 0.42\\
&& B &  0.78  & 0.42\\
&& C &  0.78  & 0.42\\
&& D &  0.59  & 0.46\\
\hline
& $2.0 < \eta_{M} < 5.0$ 
 & A & 5.58 & 0.42\\
&& B & 1.81 & 0.43\\
&& C & 1.81 & 0.43\\
&& D & 1.42 & 0.46\\
\hline
\hline
\end{tabular}
\end{table}

\section{Conclusions}
\label{conclusions}

We have discussed CEP of $\eta$ and $\eta'$ mesons 
in proton-proton collisions.
Data exist from the WA102 experiment at $\sqrt{s} = 29.1$~GeV and,
hopefully, will be collected in the future by LHC experiments
at $\sqrt{s} \approx 13$~TeV.
Our main points are as follows.

\begin{itemize}
\item[(i)]
Observation of CEP of $\eta$ and/or $\eta'$,
and/or $f_{1}$ at the LHC would \underline{not be compatible}
with a scalar character of the pomeron.
According to the tensor-pomeron model \cite{Ewerz:2013kda}
all these mesons can be produced in CEP by double-pomeron exchange.

\item[(ii)]
Comparison of CEP of $\eta$, $\eta'$ at WA102
and LHC will allow us to determine
the strengths of $\Pom \Pom$ and of non-leading Reggeon exchanges, $\Pom f_{2 \Reg} + f_{2 \Reg} \Pom$
and $f_{2 \Reg} f_{2 \Reg}$.
Here we assume that the effective couplings of $\eta$
and $\eta'$ to pomerons and $f_{2 \Reg}$ are energy independent.
This has some support from the fact that
with a constant $\Pom pp$ coupling function
$\beta_{\Pom NN}$ (\ref{2.7}) one finds a good description of $pp$ elastic scattering.
For a review see e.g. \cite{Donnachie:2002en}.
The energy dependence of the cross sections resides
in the Regge factors.
In the tensor-pomeron model these are incorporated
in the effective pomeron and $f_{2 \Reg}$ propagators;
see \cite{Ewerz:2013kda}.

\item[(iii)]
In the Table~\ref{table:1} we give cross sections to be
expected in LHC experiments.

\item[(iv)]
Sometimes one hears the following argument.
In strict flavour SU(3)$_{\rm F}$ symmetry 
the $\eta$ is an SU(3)$_{\rm F}$ octet,
the pomeron an SU(3)$_{\rm F}$ singlet.
Therefore, the fusion reaction
\begin{eqnarray}
\Pom + \Pom \to \eta
\label{4.1}
\end{eqnarray}
should be forbidden. We do \underline{not} share this argument.
Flavour SU(3)$_{\rm F}$ is badly broken and the $\eta$ meson
has a gluon-gluon component in its wave function,
with which it certainly can couple to pomerons.
Observation of CEP of $\eta$ at the LHC will definitely
clarify also this issue.

\end{itemize}

In Refs.~\cite{Lebiedowicz:2013ika,Lebiedowicz:2025num}, 
one can find details of the calculations,
many other results not discussed here,
and many references.
For other discussions of CEP in the Regge approach we refer,
for instance, to \cite{Donnachie:2002en,Kaidalov:2003fw}.

\acknowledgments
The authors would like to thank the organisers of the ISMD conference held in Corfu (Greece) in September 2025
for giving us the opportunity to present our work over ZOOM.


\end{document}